\documentclass{svproc}
\usepackage{graphicx}%
\usepackage{multirow}%
\usepackage{amsmath,amssymb,amsfonts}%
\usepackage{mathrsfs}%
\usepackage[title]{appendix}%
\usepackage{xcolor}%
\usepackage{textcomp}%
\usepackage{manyfoot}%
\usepackage{booktabs}%
\usepackage{algorithm}%
\usepackage{algorithmicx}%
\usepackage{algpseudocode}%
\usepackage{listings}%
\usepackage{url}

\usepackage{subcaption}

\usepackage{CJKutf8}

\usepackage{hyperref}

\begin{document}
\mainmatter              
\title{Mapping Hong Kong's Financial Ecosystem: \\A Network Analysis of the SFC's Licensed Professionals and Institutions}
\titlerunning{Network Analysis of HK SFC Register}  
%
\author{Abdulla AlKetbi\inst{1,2,3} \and Gautier Marti\inst{3} \and Khaled AlNuaimi\inst{1,3} \and Raed Jaradat\inst{1} \and Andreas Henschel\inst{1}}
\authorrunning{Abdulla AlKetbi et al.} 

\institute{Khalifa University, Shakhbout Bin Sultan St - Hadbat Al Za'faranah - Abu Dhabi,\\
\email{100061314@ku.ac.ae},\\ WWW home page:
\texttt{https://www.ku.ac.ae}
\and
ADIA Lab, Level 26, Al Khatem Tower, ADGM Square, Al Maryah Island,\\
Abu Dhabi, United Arab Emirates
\and
Abu Dhabi Investment Authority (ADIA)
}

\maketitle              

\begin{abstract}

We present the first study of the Public Register of Licensed Persons and Registered Institutions maintained by the Hong Kong Securities and Futures Commission (SFC) through the lens of complex network analysis. This dataset, spanning 21 years with daily granularity, provides a unique view of the evolving social network between licensed professionals and their affiliated firms in Hong Kong's financial sector. Leveraging large language models, we classify firms (e.g., asset managers, banks) and infer the likely nationality and gender of employees based on their names. This application enhances the dataset by adding rich demographic and organizational context, enabling more precise network analysis. Our preliminary findings reveal key structural features, offering new insights into the dynamics of Hong Kong's financial landscape. We release the structured dataset to enable further research, establishing a foundation for future studies that may inform recruitment strategies, policy-making, and risk management in the financial industry.

\end{abstract}

\section{Introduction}\label{sec1}

According to the Hong Kong Monetary Authority (HKMA), the financial services sector is a cornerstone of Hong Kong's economy, contributing 23\% of GDP and employing approximately 277,000 people in 2022, representing 7.6\% of total employment \cite{hkma_ifc_website}. This sector supports Hong Kong's status as a leading international financial center, with a robust infrastructure that includes retail and investment banking, asset management, insurance, and securities trading, all under the critical regulatory oversight of the HKMA and the Securities and Futures Commission (SFC).

While macroeconomic indicators like GDP and employment data offer valuable insights, they often lag behind real-time developments (e.g., the HKMA Annual Report 2023 \cite{hkma_ar_2023} published on 30 April 2024). To accurately monitor the financial sector, a more granular understanding of its dynamics is crucial. Key aspects, such as firm creation rates, the longevity of financial entities, and career trajectories within this ecosystem, provide deeper insights into the sector's health and can serve as early indicators of broader economic trends.

By applying complex network theory to the Securities and Futures Commission (SFC) \textit{Public Register of Licensed Persons and Registered Institutions}, we aim to uncover the intricate interactions between firms and employees. This bottom-up approach, informed by econophysics and complexity economics, allows us to explore how individual behaviors shape the broader economic landscape. Furthermore, by enriching our dataset with classifications derived from large language models, we can monitor the sector's cultural diversity and demographic shifts, particularly the participation of expatriates. This analysis tracks whether there is an increasing presence of professionals from Western Europe, the UK, and the US, or if there is a rising influx from other parts of Asia, especially Mainland China. Such insights provide valuable understanding of the evolving demographics of Hong Kong's financial sector and its broader economic implications.

\section{Dataset Description}\label{sec2}

\subsection{Origin of the Dataset}\label{origin}

This dataset originates from the public register maintained by the Hong Kong Securities and Futures Commission (SFC)\footnote{\url{https://www.sfc.hk/en/Regulatory-functions/Intermediaries/Licensing/Register-of-licensed-persons-and-registered-institutions}}, which has systematically recorded licensed individuals, corporations, and registered institutions since the implementation of the Securities and Futures Ordinance (SFO) on 1 April 2003. Under the SFO framework, the SFC's guidelines mandate that individuals engaging in regulated activities, such as dealing in securities or providing financial advice, must obtain a license. Conversely, individuals in non-regulated roles, such as administrative or support staff, are generally not required to be licensed. The dataset primarily encompasses roles that require high qualifications and expertise, carry significant responsibilities, command higher earnings, and play a critical role in the financial sector.

\sloppy
Additionally, the dataset includes information on virtual asset service providers (VASPs) licensed under the Anti-Money Laundering and Counter-Terrorist Financing Ordinance (AMLO) since 1 June 2023, reflecting Hong Kong's expanding regulatory scope in response to evolving financial technologies.

\subsection{Dataset Acquisition and Overview}\label{overview}

We compiled the dataset through a systematic web scraping process from the Hong Kong Securities and Futures Commission (SFC) public register. Although this register provides detailed information on licensed individuals and corporations, it is not directly accessible in a downloadable format, requiring users to query specific keywords or filter through criteria to access segments of the data.

To construct a comprehensive dataset suitable for research, we extracted all available data points related to licensees. The final dataset comprises 519,860 rows and 12 columns, capturing all relevant information on licensed persons and registered institutions under the SFO and AMLO frameworks. We have made the dataset publicly available online to facilitate further research\footnote{\url{https://www.kaggle.com/datasets/gautiermarti/hk-sfc-register}}.

A summary of the dataset's statistics is presented in \hyperref[tab:tab1]{Table~\ref*{tab:tab1}}. This table provides an overview of key data fields, including the total count of entries, the number of unique values, the most frequent entry ("top"), and the frequency of the most common entry ("freq") for each field.

\begin{table}[h]
\centering
\caption{Summary Statistics of the SFC Licensees Dataset}\label{tab:tab1}%
\begin{CJK}{UTF8}{bsmi}
\begin{tabular}{@{}lcrrr@{}}
\toprule
Data field & count  & unique & top & freq \\
\midrule
effectiveDate    & 519860   & 5395  & 2003-04-01  & 84136  \\
endDate    & 434767   & 7585  & 2008-09-29  & 1024  \\
fullname   & 519860   & 117232  & ZHANG Fan 張帆 & 93  \\
sfcid    & 519860   & 121883  & AZN097  & 65  \\
lcRole    & 519860   & 2  & RE  & 443216  \\
prinCeName    & 519860   & 4971 & Morgan Stanley Asia Limited  & 11784  \\
prinCeNameChin    & 419845   & 3558  & 摩根士丹利亞洲有限公司  & 11784  \\
prinCeRef    & 519860   & 4979  & AAD291  & 11784  \\
regulatedActivity.status    & 519860   & 2  & R  & 468255  \\
regulatedActivity.actType    & 519860   & 10  & 1  & 181390  \\
regulatedActivity.actDesc   & 519860   & 10  & Dealing in Securities  & 181390  \\
regulatedActivity.cactDesc    & 519860   & 10  & 證券交易  & 181390  \\
\bottomrule
\end{tabular}
\end{CJK}
\end{table}


To aid in understanding the dataset and its analysis, we provide brief descriptions of each data field below:

\begin{description}
  \item[\textbf{effectiveDate}] Start date of the license or regulated activity.
  \item[\textbf{endDate}] Termination or expiration date of the license or activity.
  \item[\textbf{fullname}] Full legal name of the license holder (given and family names).
  \item[\textbf{sfcid}] Unique ID assigned by the SFC to identify each licensee.
  \item[\textbf{lcRole}] Licensee’s role within the SFC framework:
  \begin{itemize}
  \item \textbf{RE}: Representative authorized to carry out regulated activities under supervision.
  \item \textbf{RO}: Responsible Officer, authorized to supervise regulated activities.
  \end{itemize}
  \item[\textbf{prinCeName}] Official English name of the firm employing the licensee.
  \item[\textbf{prinCeNameChin}] Official Chinese name of the firm.
  \item[\textbf{prinCeRef}] Unique ID assigned by the SFC to each licensed firm.
  \item[\textbf{regulatedActivity.status}] Current status of the regulated activity:
  \begin{itemize}
      \item \textbf{R}: Registered/Active.
      \item \textbf{A}: Archived/Inactive.
  \end{itemize}
  \item[\textbf{regulatedActivity.actType}] Numerical code for the type of regulated activity (e.g., \textbf{1}: Dealing in Securities; \textbf{2}: Dealing in Futures Contracts; \textbf{3}: Leveraged Foreign Exchange Trading; etc.).
  \item[\textbf{regulatedActivity.actDesc}] Description of the regulated activity in English.
  \item[\textbf{regulatedActivity.cactDesc}] Corresponding description in Chinese.
\end{description}

\subsection{Exploratory Data Analysis}\label{eda}

Having established the dataset's origin and structure, we now turn to an exploratory data analysis to uncover key patterns and trends within Hong Kong's financial sector. This analysis will provide a foundation for the subsequent network constructions.

\subsubsection{Key Statistics}

To gain an initial understanding of the dataset, key statistics were calculated and summarized in \hyperref[tab:dataset_overview]{Table~\ref*{tab:dataset_overview}}. The dataset spans more than two decades, covering the period from April 2003 to March 2024, and includes a total of 519,860 licenses issued to 121,883 employees across 4,979 firms. The relatively short median tenure of 1.5 years suggests a high turnover rate, which is characteristic of competitive financial hubs. Of the 1,336 firms active in April 2003, only 597 have maintained continuous activity to this day. Out of the 4,979 firms, 1,619 have ceased operations, with a median lifespan of 4.1 years, highlighting the volatile nature of Hong Kong's financial industry.

\begin{table}[h]
\centering
\caption{Key Statistics of the Dataset}\label{tab:dataset_overview}
\begin{tabular}{l r}
\toprule
\textbf{Statistic} & \textbf{Value} \\
\midrule
\textbf{Dataset Coverage} & \\
\midrule
Start Date & 2003-04-01 \\
End Date & 2024-03-01 \\
Total Number of Licenses & 519,860 \\
\midrule
\textbf{Employee Statistics} & \\
\midrule
Total Number of Employees & 121,883 \\
Median License Tenure (years) & 1.5 \\
Maximum License Tenure (years) & 21 \\
Median Number of Active Employees & 35,000 \\
Professionals Active Since Start Date & 5,143 \\
Average Number of Firms per Employee & 2.1 \\
Median Proportion of RO Licensees & 16\% \\
\midrule
\textbf{Firm Statistics} & \\
\midrule
Total Number of Firms & 4,979 \\
Number of Firms Active At Start Date & 1336 \\
Number of Firms Active Since Start Date & 597 \\
Number of Firms with Ceased Activity & 1619 \\
Median Lifespan of Ceased Firms (years) & 4.1 \\
Average Number of Licensees per Firm & 21 \\
\bottomrule
\end{tabular}
\end{table}

\subsubsection{License Types and Professional Specializations}

The analysis of license types reveals that professionals in the SFC register typically hold four or fewer license types. Notably, 30\% of professionals hold exactly one license type, and another 30\% hold two. The most common single license type is ``Dealing in Securities", followed by ``Advising on Securities" and ``Asset Management" (cf.~\hyperref[fig:distribution_licenses]{Figure~\ref*{fig:distribution_licenses}}). When examining combinations of license types, the most common pairings involve ``Advising on Securities" and ``Dealing in Securities". License triplets further highlight the combination of trading and markets expertise (``Dealing in Securities", ``Advising on Securities") with asset management or corporate finance (investment banking) roles.

This analysis reveals that licensees in Hong Kong's financial sector exhibit a high degree of specialization. The majority are engaged in trading activities, primarily dealing in securities, with a smaller segment focused on trading derivatives such as futures contracts. When holding multiple licenses, trading activities are often paired with licenses in ``Asset Management" (encompassing mutual funds, pension funds, hedge funds, and asset management divisions of banks and insurers) or ``Advising on Corporate Finance" (covering investment banking activities like M\&As and IPOs). These areas constitute the core of licensed activities in Hong Kong's financial industry.

\subsubsection{Job Market Dynamics: License Creations and Terminations}

\begin{figure}[htbp]
    \centering
    \begin{subfigure}[b]{0.49\textwidth}
        \centering
        \includegraphics[width=\textwidth]{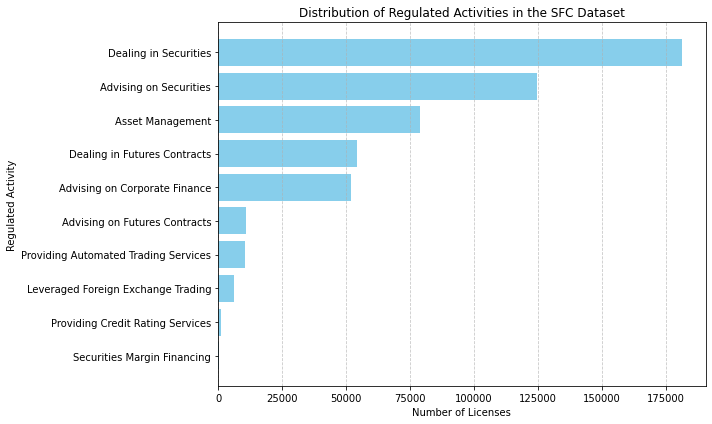}
        \caption{Distribution of regulated activities in the SFC dataset, with 'Dealing in Securities' as the most common license.}
        \label{fig:distribution_licenses}
    \end{subfigure}
    \hfill
    \begin{subfigure}[b]{0.49\textwidth}
        \centering
        \includegraphics[width=\textwidth]{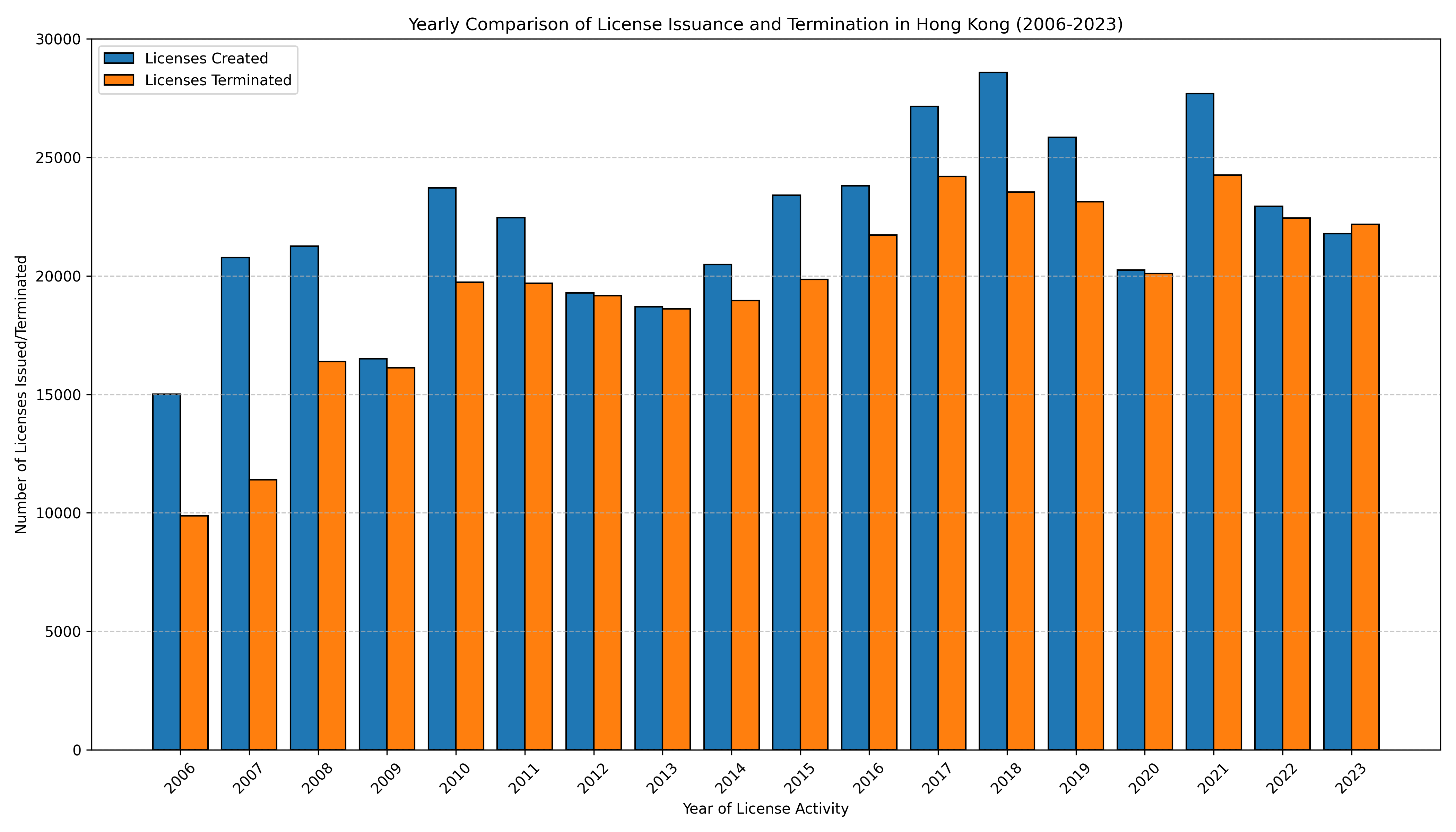}
        \caption{Yearly comparison of licenses issued (blue) vs. terminated (orange) in Hong Kong (2006-2023).}
        \label{fig:licenses_created_vs_terminated}
    \end{subfigure}
    \caption{Overview of License Distribution and Dynamics in Hong Kong's Financial Sector. The left panel shows the distribution of regulated activities, highlighting the centrality of securities trading. The right panel illustrates trends in license issuance and termination, reflecting the industry's response to major events.}
    \label{fig:combined_figures}
\end{figure}

The Global Financial Crisis (GFC) in 2009 and the onset of the COVID-19 pandemic in 2020 led to significant economic disruptions, reflected in the nearly equal numbers of terminations and creations during those years (cf.~\hyperref[fig:licenses_created_vs_terminated]{Figure~\ref*{fig:licenses_created_vs_terminated}}). The stagnation observed in 2012 and 2013, where license issuance did not outpace terminations, coincided with the introduction of stringent real estate measures aimed at cooling the overheated property market, potentially dampening financial sector expansion. In 2023, for the first time, terminations outnumbered creations, possibly signaling economic uncertainties, shifts in market dynamics, or a trend of financial services and professionals migrating to other emerging global hubs like Singapore and the UAE (Dubai, Abu Dhabi).

\subsubsection{Firm Licensing Structures and Top Entities by Licensee Count}

\hyperref[tab:license_total_count_per_firm]{Table~\ref*{tab:license_total_count_per_firm}} lists the top 20 firms in the SFC Register by the total number of licensees.

\begin{table}
\centering
\caption{Top 20 Firms by Total Number of Licensees in the SFC Register}
\label{tab:license_total_count_per_firm}
\begin{tabular}{lr}
\toprule
\textbf{prinCeName} & \textbf{Total Licensees} \\
\midrule
Morgan Stanley Asia Limited & 5527 \\
Goldman Sachs (Asia) L.L.C. & 4169 \\
OnePlatform Asset Management Limited & 4051 \\
Merrill Lynch (Asia Pacific) Limited & 3853 \\
Citigroup Global Markets Asia Limited & 3572 \\
China International Capital Corporation Hong Kong Limited & 3201 \\
Nomura International (Hong Kong) Limited & 3030 \\
Manulife Investment Management (Hong Kong) Limited & 2734 \\
Credit Suisse (Hong Kong) Limited & 2668 \\
Merrill Lynch Far East Limited & 2595 \\
UBS Securities Asia Limited & 2494 \\
Everbright Securities Investment Services (HK) Limited & 2370 \\
Phillip Securities (Hong Kong) Limited & 2231 \\
CLSA Limited & 2115 \\
Haitong International Securities Company Limited & 1883 \\
Morgan Stanley Hong Kong Securities Limited & 1762 \\
Macquarie Capital Limited & 1707 \\
AIA Wealth Management Company Limited & 1687 \\
KGI Asia Limited & 1667 \\
Phillip Commodities (HK) Limited & 1490 \\
\bottomrule
\end{tabular}
\end{table}

HSBC, despite being one of Hong Kong's leading financial institutions, doesn't appear in the top rankings by total licensees because its operations are spread across 21 entities, each handling specific activities as required by the SFC's regulatory framework. Collectively, these entities account for 4,383 licensees, placing HSBC among the top firms. This structure is common among global giants like Morgan Stanley and Goldman Sachs, which also operate through multiple entities to comply with the SFC's licensing requirements.

\subsubsection{Further Opportunities for Exploration}

Due to space constraints, we will limit our discussion here. However, potential avenues for further investigation include longitudinal trends in firm growth, employee mobility and career paths, developing firm-specific risk profiles based on employee turnover rates, shifts in regulatory frameworks, and the impact of external economic events on licensing patterns.

\subsection{Data Enrichment using Large Language Models}\label{llms}

This paper does not aim to delve into the technical intricacies of Large Language Models (LLMs) and their applications for enriching datasets. Instead, we leverage LLMs for their capacity to extract valuable features from data, drawing on their extensive memorization of world knowledge and their ability to learn complex patterns. These capabilities allow us to efficiently map names—both of individuals and firms—to attributes such as likely country of origin, gender, and business classifications, enabling a more nuanced analysis of demographic and organizational trends within Hong Kong's financial industry.

While the classification of ethnicity from names has been studied using traditional statistical methods in fields like public health, population studies and social media analysis \cite{mateos2007review,ambekar2009name,hofstra2018predicting,petersen2021names}, our approach innovates by employing LLMs, which we find to outperform these earlier techniques. We provide more details into the biases, shortcomings and opportunities of enriching datasets with demographics through the use of LLMs in \cite{alnuaimi2024enriching}.

\subsubsection{Country of Origin} 

\begin{figure}[h!]
\centering
\begin{minipage}{0.48\textwidth}
    \centering
    \captionof{table}{Top 10 Countries of Origin}\label{table:coo}
    \begin{tabular}{l r r}
    \toprule
    \textbf{Country of Origin} & \textbf{Count} & \textbf{Ratio (\%)} \\ \midrule
    China            & 80,549 & 66\\
    Hong Kong        & 8,286  & 7\\
    United States    & 4,474  & 4\\
    South Korea      & 4,078  & 3\\
    United Kingdom   & 3,959  & 3\\
    Taiwan           & 3,461  & 3\\
    India            & 3,240  & 3\\
    Japan            & 2,790  & 2\\
    Singapore        & 2,121  & 2\\
    France           & 1,575  & 1\\
    \bottomrule
    \end{tabular}
\end{minipage}\hfill
\begin{minipage}{0.48\textwidth}
    \centering
    \includegraphics[width=\textwidth]{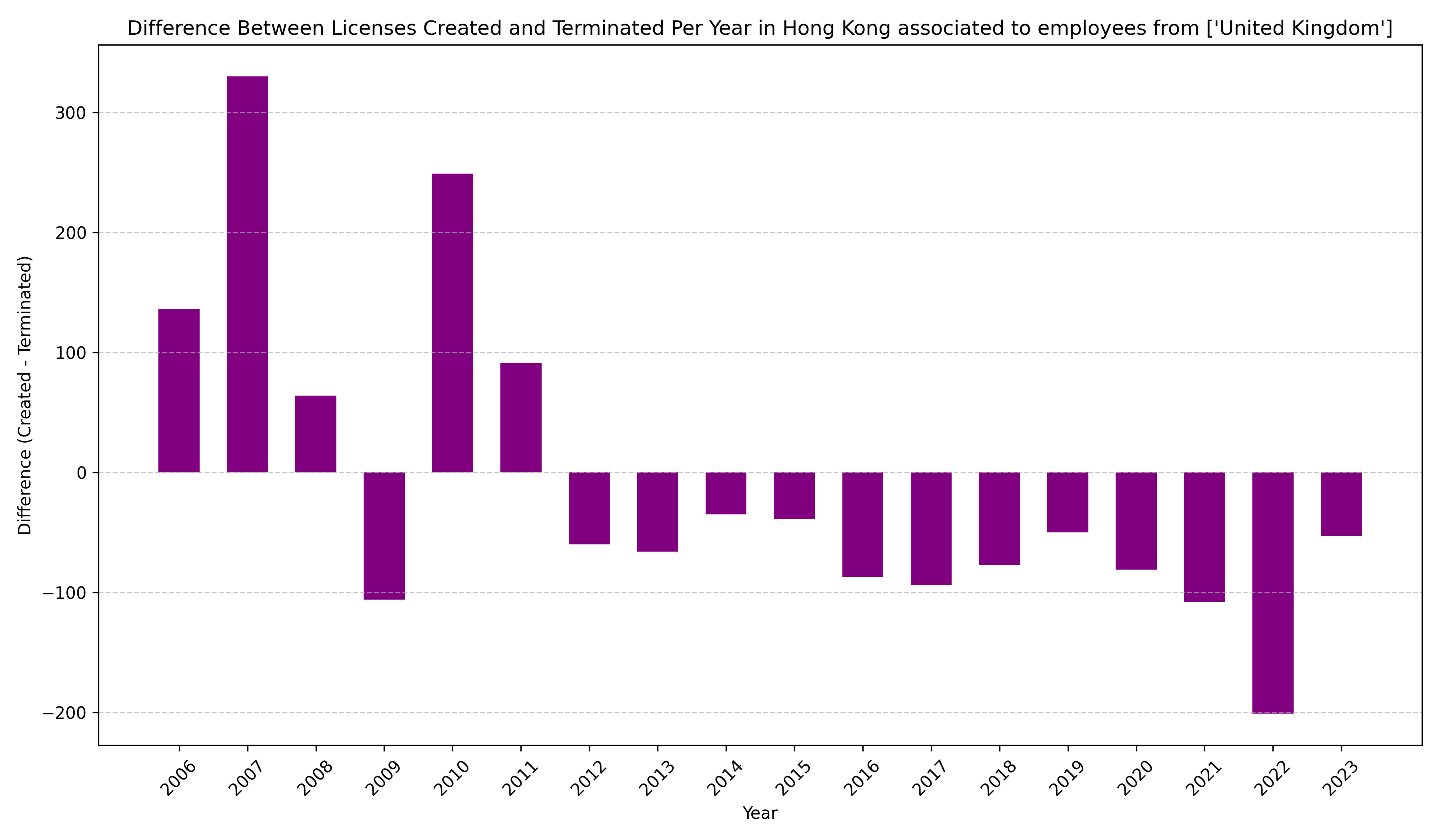}
    \caption{Licenses Issued vs. Terminated in Hong Kong for UK expatriates}\label{fig:licenses_hk_uk}
\end{minipage}
\end{figure}

We employ LLMs to estimate the country of origin of individuals based on their full names and their association with Hong Kong's financial industry. By applying this method, we infer the likely country of origin for all 121,833 individuals in the database (cf.~\hyperref[table:coo]{Table~\ref*{table:coo}}). This inferred origin serves as a proxy for nationality and background, enabling us to differentiate between local and expatriate populations.

Our findings indicate significant trends: Western expatriates are departing Hong Kong, while the presence of individuals from China has been increasing, particularly since 2014. Additionally, the number of professionals from India has declined since the onset of the COVID-19 pandemic. Notably, the British expatriate community has been shrinking since 2012 (cf.~\hyperref[fig:licenses_hk_uk]{Figure~\ref*{fig:licenses_hk_uk}}).

\subsubsection{Gender}

The gender distribution among licensed professionals reveals stark contrasts across different nationalities. Notably, Western expatriates (Europe, North America) exhibit the lowest representation of female professionals, with only 10\% of their licensed workforce being female. This contrasts sharply with higher female representation among Asian nationalities, particularly from Japan (20\%), South Korea (25\%), China, and Southeast Asian countries, where the proportion ranges from 30\% to 40\%. These differences may reflect cultural, societal, or industry-specific factors, warranting further investigation.

While our data enrichment process offers insights into gender distribution, it is important to consider potential biases in the LLMs used to infer gender from names. Accuracy varies by cultural context, with higher reliability for Western, Japanese, and Korean names, which are more gender-specific. In contrast, Chinese and some Southeast Asian names, often unisex or challenging to Romanize, may lead to misclassification.

\subsubsection{Buy-Side vs. Sell-Side}

The SFC public register lists affiliated firms without detailed activity descriptions. We addressed this by using state-of-the-art LLMs, such as \texttt{claude-3.5-sonnet} \cite{anthropic2024claude} (a closed-source model) and \texttt{llama3-70b} \cite{dubey2024llama} (an open-source model), to classify firms as buy-side or sell-side. Buy-side firms, including asset managers and hedge funds, focus on investment, while sell-side firms, like investment banks and broker-dealers, facilitate market transactions. Our analysis shows a significant increase in buy-side firms, rising from 25\% in 2003 to 60\% in 2024, with LLMs showing 90\% concordance in firm classification.

\subsubsection{Firm Classification}

In our analysis, we classified each firm in the dataset into one of 13 predefined categories based on their primary business activities. The resulting distribution of firms across the most frequent categories is summarized in \hyperref[table:cls]{Table~\ref*{table:cls}}.

\begin{table}
\centering
\caption{Distribution of Firms and Employees by Type (Top 5 Categories)}\label{table:cls}
\begin{tabular}{lccr}
\hline
\textbf{Company Type} & \textbf{\% total firms} & \textbf{\% total employees} & \textbf{Activity Type}\\
\hline
Asset Management Firms       & 44\% &  30\% & buy-side \\
Brokerage Firms              & 23\% & 31\% & sell-side \\
Investment Banks             & 10\% & 31\% & sell-side \\
Private Equity Firms         & 10\% & 4\% & buy-side \\
Hedge Funds                  &  5\% & 4\% & buy-side \\
\hline
\end{tabular}
\end{table}

Our findings reveal that buy-side firms constitute a larger share of the total firms in the dataset, accounting for over 60\%. However, they employ less than 40\% of the workforce, indicating smaller-scale operations. In contrast, sell-side firms, which make up less than 33\% of the total firms, typically operate on a larger scale, employing more than 60\% of the total workforce. This distribution suggests that Hong Kong's financial landscape is characterized by a high number of specialized buy-side firms, which are likely focused on niche markets or managing smaller, targeted portfolios.


\section{Network Constructions}\label{sec3}

The dataset provides a robust foundation for constructing socio-economic networks, with two primary sets of nodes: individuals (Hong Kong finance professionals) and firms (regulated financial institutions). Various network constructions can be derived from this dataset, each suited to different analytical purposes. The choice of network structure is crucial, as it shapes the insights that can be drawn and the complexity of the systems that can be analyzed.

While an exhaustive exploration of all possible network constructions is beyond the scope of this paper, we aim to introduce the dataset and stimulate further research on the complex systems reflected in regulatory licenses within the financial industry. To illustrate the potential of this dataset, we present two fundamental network constructs: a firm-firm network and an employee-employee network. These examples demonstrate the intricate and complex nature of the relationships within Hong Kong's financial ecosystem.

\subsection{Firm-Firm Network Based on Shared Employees}

One network construction connects firms based on the employees they have shared over time. Specifically, we construct a temporal graph \( G_t = (V_t, E_t) \) for each time point \( t \), where \( V_t \) represents firms active at time \( t \), and \( E_t \) represents edges between firms that have employed the same individuals.

\subsubsection{Vertices} 
The vertex set \( V_t \) consists of firms that have at least one active employee at time \( t \): $
V_t = \{ v_i \in \text{firms} \mid \exists \, \text{employee} \, e_j \text{ such that } \text{effectiveDate}(e_j) < t \text{ and } (\text{endDate}(e_j) > t) \}.$

\subsubsection{Edges} 
An edge \( (v_i, v_j) \) exists between two firms \( v_i \) and \( v_j \) at time \( t \) if they share at least one employee who worked at both firms prior to time \( t \):
$$
E_t = \{ (v_i, v_j) \mid v_i, v_j \in V_t, \, v_i \neq v_j, \, \text{ and } | \text{Employees}(v_i) \cap \text{Employees}(v_j) | > 0 \}.$$

\subsubsection{Edge Weights}
Each edge \( (v_i, v_j) \) is weighted by the number of shared employees between firms \( v_i \) and \( v_j \):
\[
w_{ij} = | \text{Employees}(v_i) \cap \text{Employees}(v_j) |
\]
We also introduce a normalized edge weight to account for the relative size of the firms and their employee turnover:
\[
w'_{ij} = \frac{| \text{Employees}(v_i) \cap \text{Employees}(v_j) |}{\sqrt{| \text{Employees}(v_i) | | \text{Employees}(v_j) |}}
\]

\subsubsection{Network Properties} We analyzed key network properties—degree distribution, average path length, and clustering coefficient—to explore the structural complexity of Hong Kong’s financial ecosystem.

Comparisons with the Erdős-Rényi random graph model \cite{erdos1959random}, a standard baseline in network analysis, were conducted to benchmark these properties. The model generates networks by randomly connecting nodes with a fixed probability, serving as a neutral reference point to highlight the unique structural features of the actual network. As of February 15, 2024, the firm-firm network comprised 3,116 nodes (firms) and 95,516 edges (connections via shared employees), forming a nearly connected component with only two isolated firms ('Bentley Capital Limited' and 'Bentley Reid \& Company Limited'), which had just eight employees over the past 21 years.

\textit{Degree Distribution:} The network exhibits a heavy-tailed degree distribution, indicative of a scale-free structure, where a few firms act as major hubs within the industry. These hubs are highly interconnected, employing a large number of professionals who have worked across multiple firms. This contrasts with the more localized distribution observed in the Erdős-Rényi model, as shown in \hyperref[fig:degree_distribution]{Figure~\ref*{fig:degree_distribution}}.

\textit{Average Path Length:} The network’s average path length is 2.5, exceeding the 1.95 observed in the Erdős-Rényi model, suggesting the presence of distinct clusters within the network.

\textit{Clustering Coefficient:} The real network has a significantly higher clustering coefficient (0.42) than the random graph (0.02), indicating the formation of tightly-knit groups or communities among firms.

\textit{Community Structure:} Clusters identified via the Louvain algorithm \cite{blondel2008fast} appear to correlate with firm types. Further analysis is required to fully understand the composition and significance of these communities.

\subsection{Employee-Employee Network Based on Shared Employers}

Another network construction focuses on the relationships between employees based on their shared employment history. Here, \( G_t = (V_t, E_t) \) is a network where each vertex \( v \in V_t \) represents an employee active at time \( t \), and each edge \( e = (v_i, v_j) \) indicates that employees \( v_i \) and \( v_j \) have worked together at the same firm.

\subsubsection{Edge Weights} 
The weight of an edge \( (v_i, v_j, w_{ij}) \) is determined by the total number of days that employees \( v_i \) and \( v_j \) have overlapped in their employment at any firm:
\[
w_{ij} = \text{OverlapDays}(v_i, v_j)
\]
We also define a normalized weight to account for the total duration of each employee's career:
\[
w'_{ij} = \frac{\text{OverlapDays}(v_i, v_j)}{\max(\text{totalDuration}(v_i), \text{totalDuration}(v_j))}
\]

\subsubsection{Network Properties}

The employee-employee network, like the firm-firm network, exhibits a high degree of connectivity, with 99\% of nodes residing in a single connected component. The network's average path length is 2.8, notably longer than the 1.7 average path length of an equivalent Erdős-Rényi random graph. The degree distribution is heavy-tailed, in stark contrast to the random graph's more localized degree distribution around 412. Additionally, the clustering coefficient of the employee-employee network is significantly higher at 0.7, compared to 0.01 for the random graph.

\begin{figure}[htbp]
    \centering
    \begin{subfigure}[b]{0.45\textwidth}
        \centering
        \includegraphics[width=\textwidth]{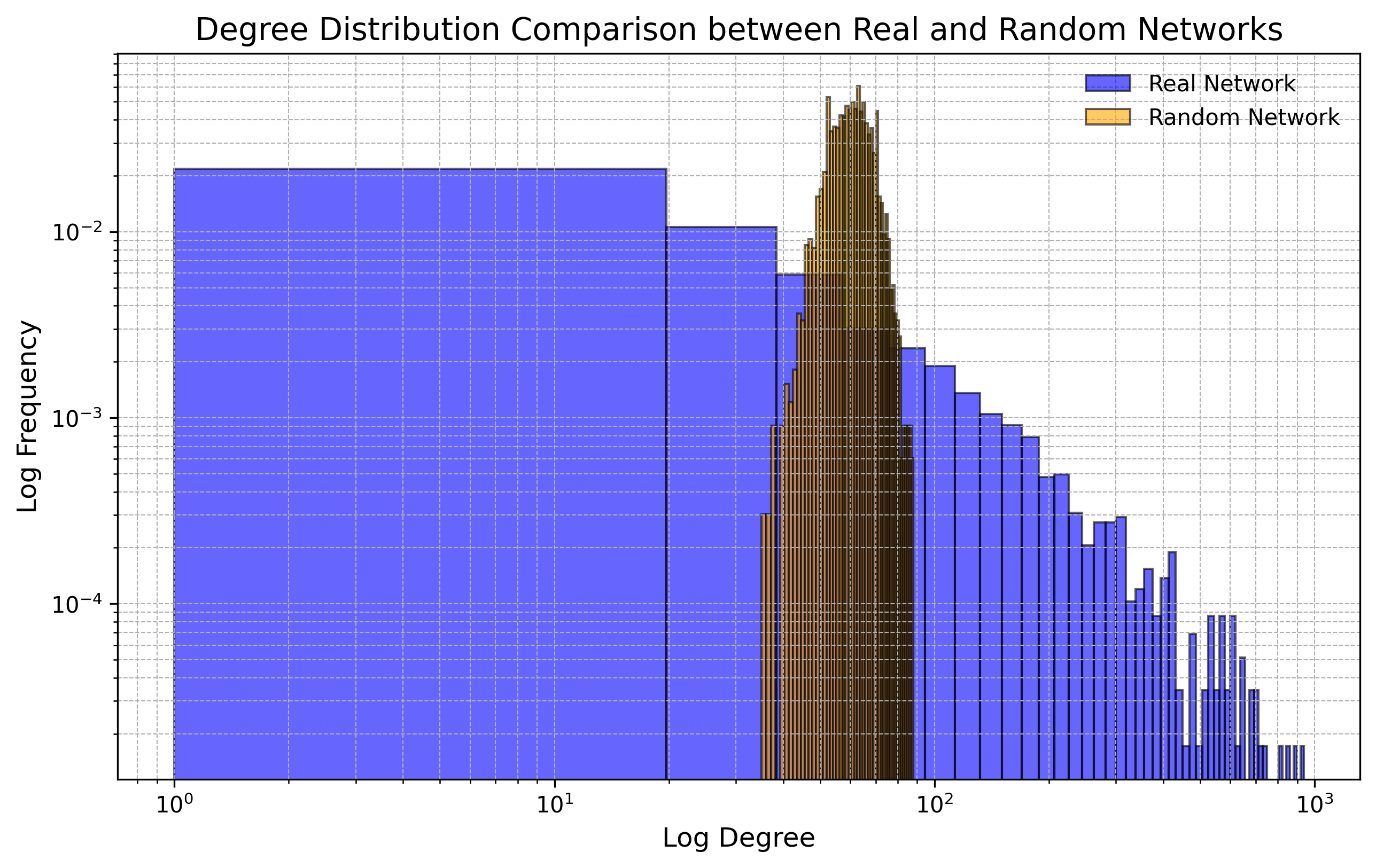}
        \caption{Degree Distribution Comparison}
        \label{fig:degree_distribution}
    \end{subfigure}
    \hfill
    \begin{subfigure}[b]{0.45\textwidth}
        \centering
        \includegraphics[width=\textwidth]{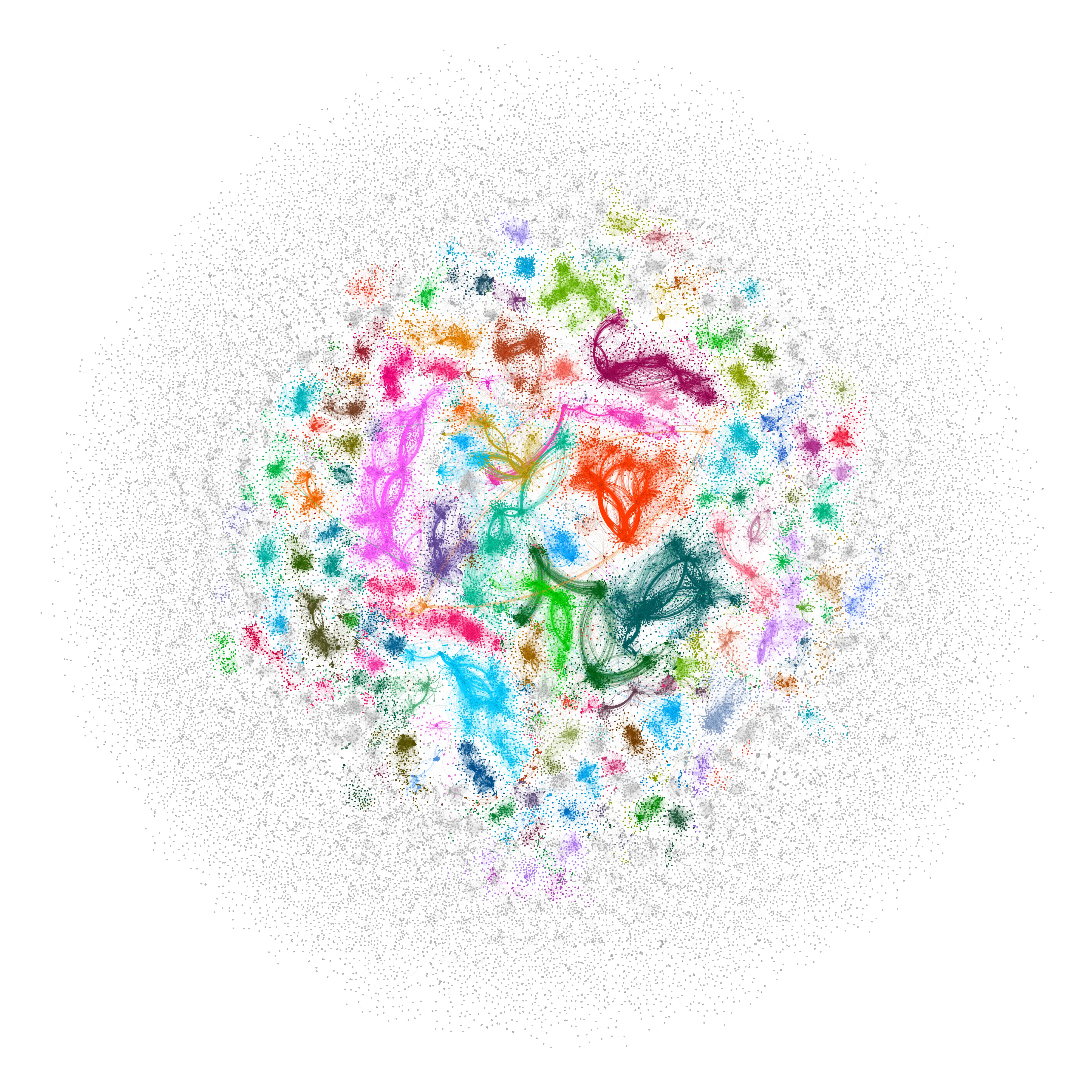}
        \caption{Employee Network Clustering}
        \label{fig:employee_clustering}
    \end{subfigure}
    \caption{Key Network Properties of Hong Kong's Financial Ecosystem: (a) Degree distribution comparison between real and random networks; (b) Visualization of employee network clustering.}
    \label{fig:network_properties}
\end{figure}

Community detection using the Louvain algorithm \cite{blondel2008fast} reveals distinct clusters within the network, as illustrated in \hyperref[fig:employee_clustering]{Figure~\ref*{fig:employee_clustering}}.

\section{Conclusion}\label{sec6}

In this study, we explored the structure of Hong Kong's financial ecosystem through network analysis of the SFC's Public Register, enriched with classifications from Large Language Models (LLMs). Our analysis revealed key structural properties, such as the heavy-tailed degree distribution and high clustering in both firm-firm and employee-employee networks, indicating a complex and interconnected financial landscape. The release of this structured dataset is a significant contribution, providing a valuable resource for future research into the dynamics of financial networks. This work opens up avenues for further studies, including the development of predictive models that leverage network structures to improve forecasting accuracy (e.g., economic variables, employee turnover, or firm-specific risks). Future research could extend this analysis to track global firm activities, explore cross-regional employee movements (exploiting registers from the Monetary Authority of Singapore (MAS) and the Financial Conduct Authority (FCA) in the UK), and examine the implications of these complex networks on financial stability. By laying the groundwork for advanced network analysis, this study contributes to a deeper understanding of financial ecosystems, offering tools and insights that can shape future research, policy-making, and risk management in the global financial industry.

\bibliographystyle{spmpsci} 

\end{document}